\documentstyle[emulateapj,epsf]{article}
\def\l{\ell}

\def\w{{\bf w}}
\def\x{{\bf x}}
\def\NN{{\bf\Sigma}}
\def\SS{{\bf S}}
\def\etal{{\frenchspacing et al.}}

\slugcomment{Submitted to Astrophysical Journal Letters August 5, 1998}

\lefthead{Devlin et al.}
\righthead{MAPPING THE CMB I}

\begin{document}

\title{MAPPING THE CMB I:  THE FIRST FLIGHT OF THE QMAP
EXPERIMENT} 

\author{M. J. Devlin\altaffilmark{1,2}, 
A. de Oliveira-Costa\altaffilmark{2,3},
T. Herbig\altaffilmark{2,4}, 
A. D. Miller\altaffilmark{2},
C. B. Netterfield\altaffilmark{5,2}, 
L. Page\altaffilmark{2} \&
M. Tegmark\altaffilmark{3,4}}

\altaffiltext{1}{Dept. of Physics, University of Pennsylvania, Philadelphia, PA 19104} 
\altaffiltext{2}{Dept. of Physics, Princeton University, Princeton, NJ 08544} 
\altaffiltext{3}{Institute for Advanced Study, Olden Lane, Princeton, NJ 08540} 
\altaffiltext{4}{Hubble Fellow}
\altaffiltext{5}{California Institute of Technology, MS 59-33, Pasadena, CA 91125}


\begin{abstract}
We report on the first flight of the balloon-borne {\sl QMAP } experiment.
The experiment is designed to make a map of the cosmic microwave
background anisotropy on angular scales from $0\fdg70$ to
several degrees.
Using the map we determine the angular power spectrum of the
anisotropy in multipole bands from $\l\sim 40$ to $\l\sim 140$.  
The results are consistent with the Saskatoon (SK) measurements. 
The frequency spectral index (measured at low $\l$) 
is consistent with that of CMB and inconsistent with either Galactic 
synchrotron or free-free emission. 
The instrument,
measurement, analysis of the angular power spectrum, and possible systematic
errors are discussed.

\end{abstract}


\section{Introduction}

Measurements of the anisotropy of the cosmic microwave background (CMB)
are an effective probe of the state of the universe roughly 300,000
years after the big bang (\cite{whi94,bon96}). The anisotropy in the CMB
not only provides a
key test of models of structure formation, but analysis of its power
spectrum may determine values of fundamental cosmological parameters
such as $\Omega_0$, $H_0$, and $\Omega_b$.

The anisotropy in the CMB is measured with a variety of 
detectors and observing strategies (\cite{ben97}, \cite{pag97}). QMAP
was designed to directly  produce a
degree scale map of the sky that could be analyzed for anisotropy.
In this paper, we report on the QMAP instrument and the first of
two balloon flights. We present maps of the microwave sky and
determinations of the angular power spectrum covering scales 
between $9^\circ$ to $0\fdg7$
in two frequency bands 
centered on 31 and 42 GHz. An analysis of the
second flight, systematic effects and calibration is presented
in Herbig {\etal} (1998, hereafter H98), 
and the data analysis methods are
described in detail in de Oliveira-Costa {\etal}  (1998, hereafter dO98) 
together with results from combining the flights.

\section{Instrument}

The QMAP balloon-borne telescope was designed, built and flown twice in
a period of 20~months.  It consists of a cryogenic receiver, primary and
secondary optics, pointing system, and data acquisition system.

The receiver, which was cooled to 2.3\,K in flight, has four feed
horns; one at Ka-band (31~GHz) and two at Q-band (42~GHz) each
with 6-7~GHz of bandwidth \footnote{Though we flew a forth feed
horn with a low noise 144 GHz SIS receiver (\cite{ker93}), it did
not  work on either flight. The problem was traced to a faulty LO
in the first flight and remains undiagnosed in the second flight.
The system now works in the ground-based MAT experiment. Also,
HEMT amplifiers have improved considerably since this time
(\cite{posp97})}. Each Ka-band and Q-band band feed horn has
two HEMT-based (high electron mobility transistor) amplifiers
(\cite{posp92}),  one in each polarization. These have been fully
characterized and are described in \cite{wol97} and \cite{mon96}.
To account for the variable temperature and consequential gain
drifts of the receiver,  a thermally stabilized noise source
calibrates all HEMT channels every 100 seconds with a 54~ms
pulse. The noise power spectrum of each of the channels used in
the final analysis is shown in Figure~1. The final map, discussed
in Section 4.2, is based on the total sensitivity of the
instrument, corresponding to the inverse quadrature sum of all of the
detector noise power spectra.

\vskip-1.3cm
\centerline{\vbox{\epsfxsize=9.0cm\epsfbox{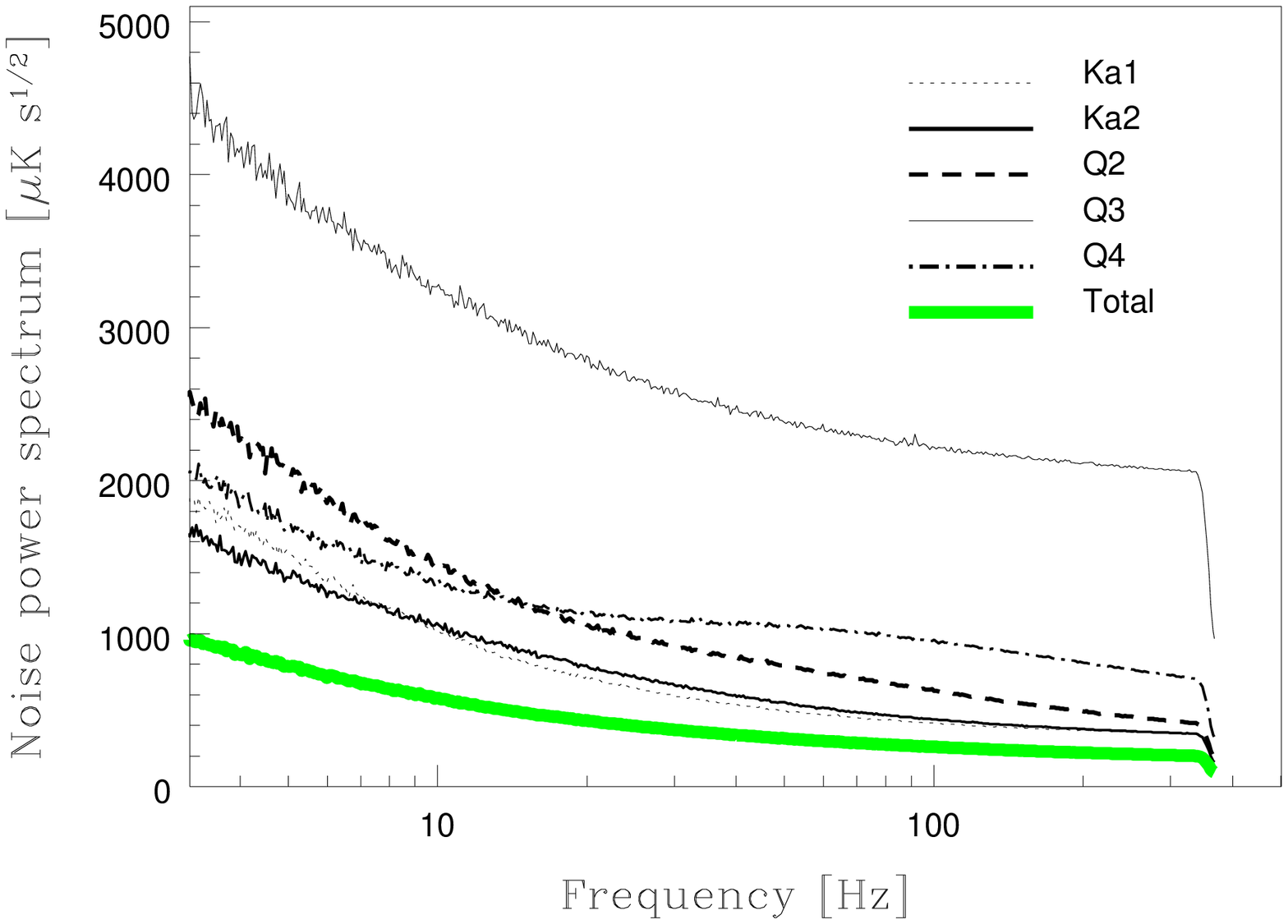}}}
\vskip-1.4cm
\figcaption{Noise power spectra of each of the channels that were used
in the final analysis.  The bottom curve is the inverse quadrature sum of all
the channels.\label{fig1}}
\medskip

The telescope optics are similar to those used for three ground
based observations in Saskatoon, SK 
(\cite{net97}; Wollack {\etal} 1997).
The telescope is mounted on an attitude-controlled  balloon platform.   
The beams are formed by cooled corrugated feeds which
under-illuminate an ambient temperature 0.85\,m off-axis parabolic
reflector which in turn under-illuminates a 1.8\,m$\times$1.2\,m chopping flat
mirror. The close-packed array of feeds form
$0\fdg89\pm0\fdg03$,   $0\fdg66\pm\fdg02$
and $0\fdg70\pm0\fdg02$ beams for Ka1/2, Q1/2, and Q3/4
respectively.  The beams are moved on the sky rapidly ($\approx
4.7$~Hz) by the computer controlled resonant
chopping flat which requires $\approx20$\,W to operate.  The
output of the detectors are AC coupled and sampled 160 times
during each chopper cycle.  This rate is intended  to adequately
oversample the sky when the beam is moving at its fastest rate.
The center of the $2\fdg7\times 2\fdg7$ array was fixed
at an elevation of $40\fdg7$ for the first flight and 
$40\fdg1$ for the second flight.   The telescope is inside a
large aluminum ground screen which is fixed with respect to the
receiver and mirrors.  

\vskip-1.3cm
\centerline{\vbox{\hglue-0.9cm\epsfxsize=10.5cm\epsfbox{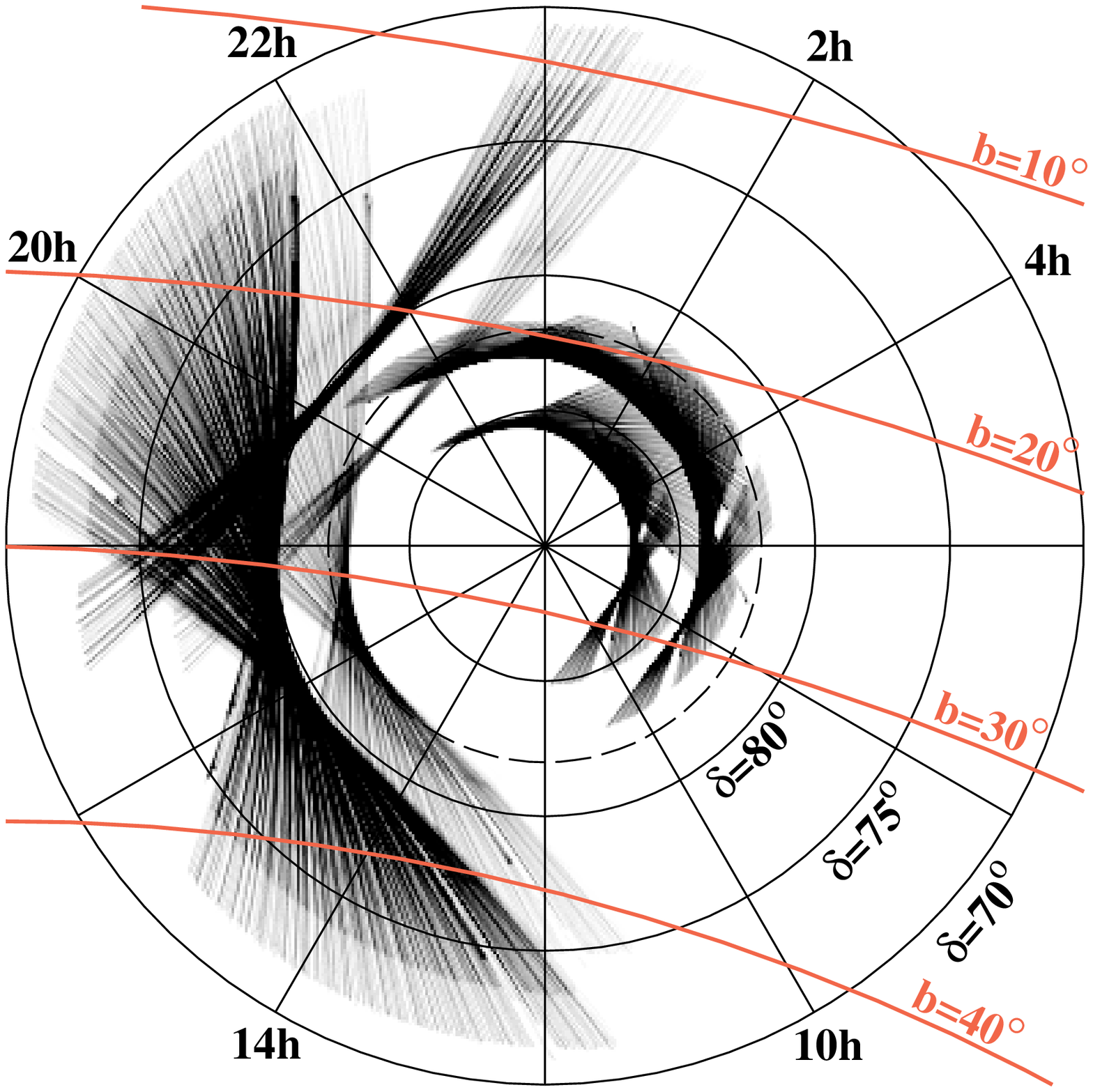}}}
\vskip-2.4cm
\figcaption{Sky coverage for flight 1 (left) and flight 2 (right) 
with intensity proportional to the time spent per pixel.  
Breaks in the coverage are due to calibrations. The scan paths
are shown for all beams. For flight 1, Q3/4 are nearer 
the NCP (center of plot).
SK observed the region inside the dashed circle.
\label{fig2}}
\medskip

The gondola is designed to be small, lightweight and rugged.
This allows us to use a smaller, more manageable balloon which increases
the probability of a successful launch and reduces the solid angle
subtended by the balloon.  The package was recovered from
both flights with virtually no damage and is currently being used on a
similar ground-based observation in Chile  
\footnote{
\label{wwwFootnote}
see
http://dept.physics.upenn.edu/cmb.html and
http://pupgg.princeton.edu/$\sim$cmb/welcome.html for more information about the
telescope in its balloon and ground-based configurations.}.  

The attitude control system obtains primary pointing information from
either a magnetometer or a CCD based star camera.  When in CCD mode, the
centroid of the chosen guide star in the CCD field is found in real
time.  This information is fed into a PID loop which controls the torque
applied to a reaction wheel at the bottom of the gondola.  Since the
Ka/Q band receivers are relatively insensitive to pendulation induced
atmospheric signals, no active elevation stabilization is required.
This straightforward pointing system allows us to achieve good observing
efficiency.  During the first flight we integrated for 4.7 hours with no
interruptions or pauses. A detailed description of the pointing system
can be found in \cite{bea96}.

\section{Observations}

The payload was flown on June 16, 1996 from Palestine, TX and November 8,
1996 from Ft.~Sumner, NM.  Using a 4~mcf balloon, observations were
made from an altitude of 30~km.  In both flights, the entire instrument
functioned nominally.

To produce high-quality maps from the data, a pixel should be referenced
to its neighbors over multiple time scales, directions, and angular
separations (\cite{wright96,teg97}).  This allows one to connect each pixel
to many others and remove inevitable instrumental drifts and offsets
in the data analysis.
Our strategy is to sweep the beams horizontally with a
$20^{\circ}$ peak-to-peak amplitude at 4.7 Hz. In addition, the entire gondola
is sinusoidally wobbled, about the meridian containing the
NCP, with a peak-to-peak amplitude of $10^{\circ}$ with a period
of 100 seconds. As the flight progresses,
one covers the sky as shown in Figure 2. In each sweep,
$\theta_{throw}/\theta_{beam}$ is 20 and 30 for the Ka and Q-band,
respectively (with $\theta_{throw}/\theta_{beam}<10$, one has
difficulty distinguishing features in the angular spectrum). A pixel is
re-observed on time scales of 0.23\,s from the chop, 100\,s from the
wobble, 12 minutes between the upper and lower beams, and $\approx 2$
hours from the rotation of the earth. With a single long flight, or
multiple shorter flights, a pixel is referenced to its neighbors
in orthogonal directions.  
The chop and wobble amplitudes were reduced by approximately
a factor of two for the second flight to increase the integration time
per pixel.  However, the basic scan strategy remained the same.

\section{Analysis}
\subsection{Data Reduction and Calibration}

For the channels used in the analysis, there are no cuts to the
data, no glitches were removed, and there is no 
detected
interference from any
on-board component \footnote{Q1 was not used in the first flight due to
interference for two hours and
Q3 lost one stage of amplification. }.
A total of 12,884,480 temperature measurements were obtained in each 
channel. 
Removal of the internal calibration every 100 seconds reduced this by 4.8\%.
A time-independent radiometric offset of 10-15~mK (peak-to-peak)
was present in all channels during the first flight. 
Much of this offset was traced to gap
in the chopper baffle.  For the second flight the offset was
1-2~mK (part of the reduction is due to a factor of two smaller chop
angle).  In both cases the offset is removed from
each channel as described in H98 and dO98. The absolute
pointing is determined during observations of Cas-A.  With the pointing 
solution, the data are binned on the sky and analyzed as a map.
The data are calibrated using Cas-A as described in H98.

\clearpage
\subsection{Map Production}

Making a map by simply averaging the observations in each 
sky direction would be inappropriate for two reasons:
scan-synchronous offsets would cause artifacts, 
and $1/f$-noise would make the map unnecessarily noisy.
We therefore adopted a more sophisticated approach, summarized below
and described in detail in dO98.

\bigskip
\medskip
\centerline{\vbox{\epsfxsize=9cm\epsfbox{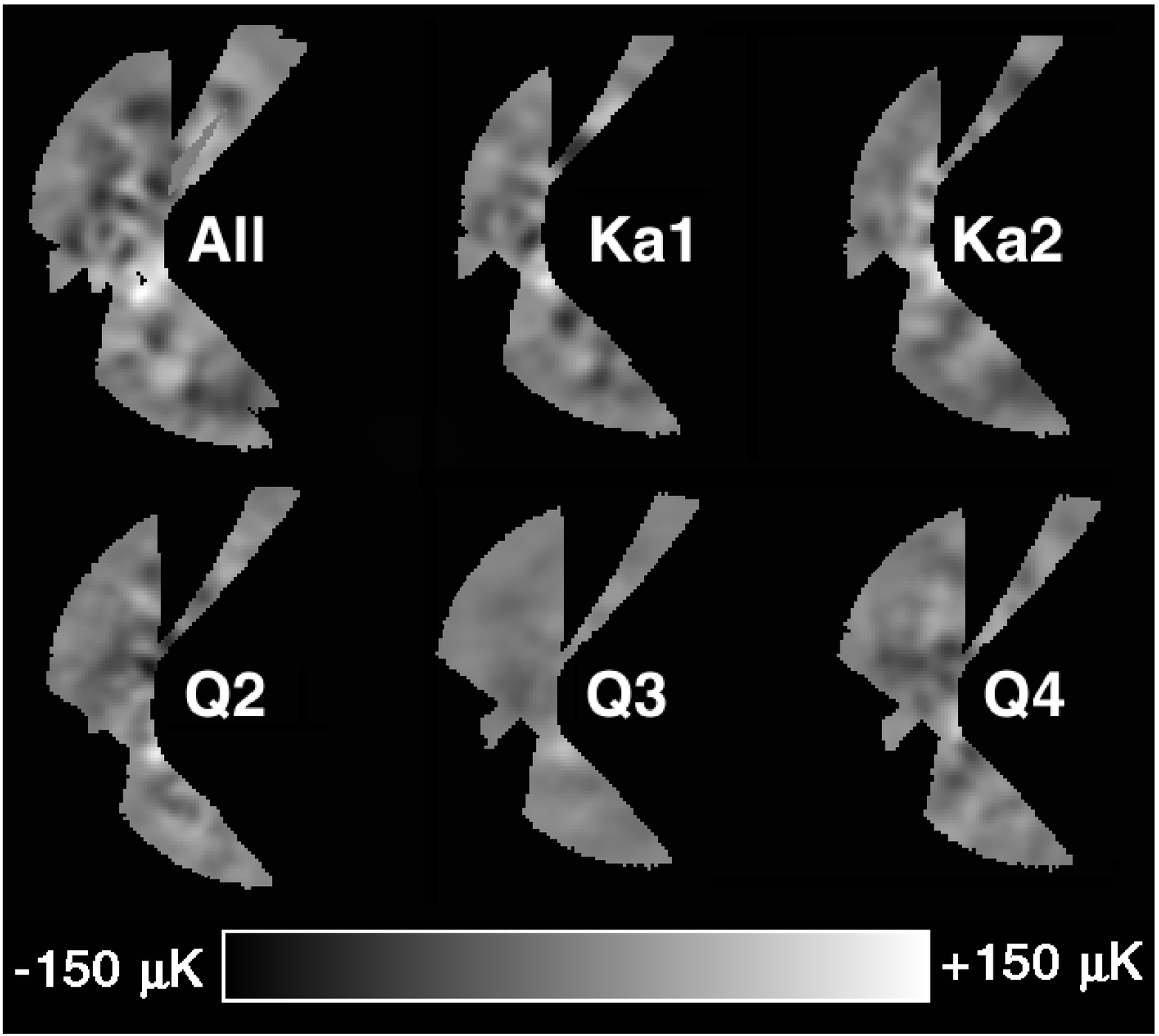}}}
\figcaption{
Wiener-filtered maps from flight 1. The CMB temperature is shown in 
coordinates where the NCP is near the center of each map, 
with RA being zero at the top and 
increasing clockwise.
}
\bigskip
\medskip

The synchronous offsets, predominantly at the chopper frequency and the
first overtone add mK-level signals to the map.  This signal is
eliminated by convolving the data with
a notch filter that vanishes at these two frequencies,
creating a data set completely independent 
of these offset harmonics and any slow variation in their amplitude.
We choose this convolution filter to eliminate the 
DC component as well, and use the remaining degrees of 
freedom in the filter to render the filtered
noise as close to white as possible. This produces a data
set where each element is the sum of three terms: 
\begin{enumerate}
\item A known linear combination of the actual sky temperatures;
\item A known linear combination of the higher-frequency synchronous
offset, which we assume is constant; and 
\item filtered detector noise, whose correlations vanish for time 
separations exceeding $\sim 10^2$ samples.
\end{enumerate}
In other words, we have a greatly overdetermined system of 
linear equations. The map we compute is the solution
to this linear inversion problem that minimizes the map noise,
as described in dO98.  For each of the five channels, 
the result of the mapmaking process is an $N$-dimensional map vector 
$\x$ (containing the temperature at each of the $N$ pixels)
and an $N\times N$-dimensional covariance matrix $\NN$ 
which characterizes the pixel noise.

Wiener filtering is a useful tool for visualizing the data. It suppresses
the noisiest modes in a map and shows signal that 
is statistically significant. 
The Wiener-filtered versions of our maps are shown in Figure 3
and are given by $\x_w\equiv\SS[\SS+\NN]^{-1}\x$, where
$\SS$ is the covariance matrix corresponding to the level of sky fluctuations 
that we find in the data.
 
\subsection{Map tests}

A generalized $\chi^2$-test for ruling out the
null hypothesis that a given map contains no signal, merely noise,
is discussed in dO98.
Applying this test to the Ka1, Ka2, Q2, Q3 and Q4 channels
shows that that signal is detected at the significance level of 
$17\sigma$, $8\sigma$, $5\sigma$, $0.4\sigma$ and $3\sigma$,
respectively.
To test whether this significant signal is common to the maps or
due to systematic errors, we apply the same test  to weighted
difference maps of the form $\x\equiv\x_1-r\x_2$ for different
relative weights $r$. The comparison of Ka1 with Ka2 is shown in
Figure 4, and it is seen that they pass three independent tests:
$\x$ is inconsistent with noise both for $r=0$ and $r=\infty$ (in
which case only Ka1 or only Ka2 are probed), but perfectly 
consistent with noise for $r=1$ (which gives the difference map
Ka1-Ka2). Likewise, we find no evidence of any signal in the
Q-band difference maps.

To see whether the signal can be accounted for by  galactic
foregrounds, we use this test to  compare the Ka-band maps to
those in the Q-band,  writing $r=(\nu_{Ka}/\nu_Q)^\beta$. As seen
in Figure 4, this places a  $2\sigma$ lower limit on the spectral
index $\beta$ of $-1.4$,  which means that the signal cannot be
explained by foregrounds such as free-free emission
($\beta\sim-2.15$) or synchrotron
radiation ($\beta\sim -2.8$) alone.  For the CMB, $\beta=0$.

\bigskip
\centerline{\vbox{\epsfxsize=9.0cm\epsfbox{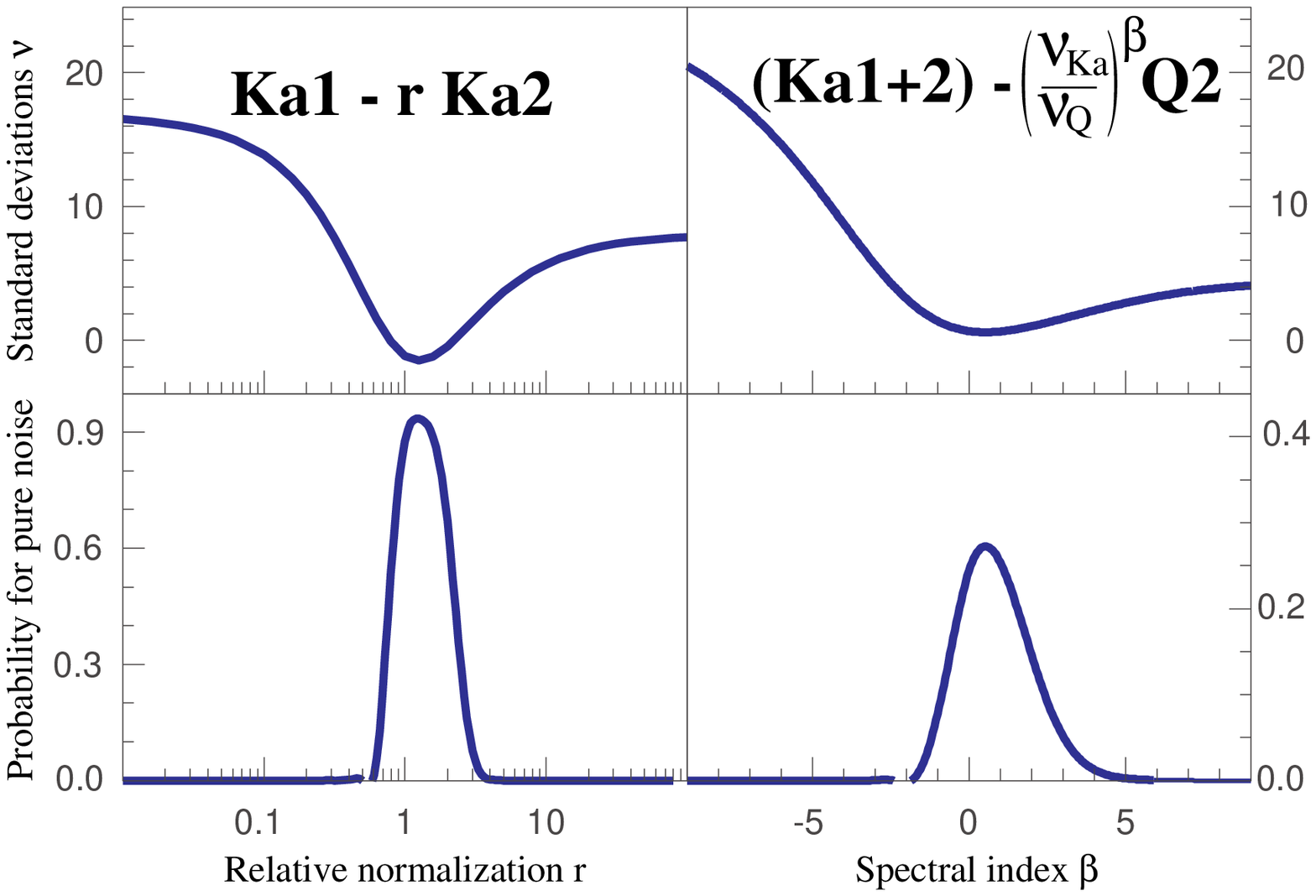}}}
\figcaption{
Evidence for common signal in the maps.
The number of $\sigma$ at which signal is detected in the 
weighted difference maps $\x_1-r\x_2$ 
is shown (top) together with the 
corresponding probability that pure noise would give such a
large $\chi^2$-value (bottom). For maps at different frequencies
(right), the spectral index $\beta =  \ln r/\ln[\nu_{Ka}/\nu_Q]$
is constrained.
}
\medskip

\subsection{Angular Power Spectrum}

We compute the angular power spectrum by expanding our maps in
signal-to-noise eigenmodes (\cite{bon95}, \cite{bun95}); 
weight vectors $\w$ that solve the generalized eigenvalue
equation $\SS\w=\lambda\NN\w$. When the $\w$ are sorted by
decreasing eigenvalue $\lambda$, they tend to probe from larger to
smaller angular scales. 
We obtain a statistically independent power
estimate from each mode, then average these individual 
estimates with inverse-variance weighting to 
obtain the band power estimates shown in Figure 5.
For more details, see dO98. 

%
%

\section{Conclusions}

The first QMAP flight has produced a 441 square degree map
of the microwave sky at an angular resolution of $\sim 0\fdg7$,
detecting sky signal that is statistically significant
at $>15\sigma$. Systematic checks confirm that the
signal is fixed on the sky and not of instrumental origin.  Spectral
analysis shows that the dominant fluctuating component is the CMB.  A
future paper will address possible levels of contamination by foreground
emission.  This measurement is consistent with the Saskatoon results
(\cite{net97}).

\acknowledgments

We gratefully acknowledge the contribution and support of many people
associated with this project. Norm Jarosik provided 
valuable insights and beautiful electronics. Dave Wilkinson and
Steve Meyer provided many useful discussions.  
Marian Pospieszalski and Mike Balister of
NRAO provided the HEMT amplifiers without which this experiment
would not have been possible. The staff at the National Scientific
Balloon Facility provided invaluable support for our flights.  We also
thank Jed Beach, Stuart
Bradley, Chris Gable, Glen Monnelly, Andrea Wood, and the
Princeton University Machine Shop for their help in constructing the
apparatus, and Suzanne Staggs for helpful comments on the manuscript.
The window functions are available at 
the web sites$^{\ref{wwwFootnote}}$, where
the calibrated raw data with pointing will be made public after
publication of this {\it Letter}.


This work was supported by
a David \& Lucile Packard Foundation Fellowship (to LP),
a Cottrell Award from Research Corporation, an NSF NYI award, 
NSF grants PHY-9222952 and PHY-9600015, 
NASA grant NAG5-6034 and Hubble Fellowships 
HF-01044.01$-$93A (to TH) and HF-01084.01$-$96A (to MT)
from by STScI, operated by AURA, Inc. under NASA contract NAS5-26555.

\vskip-1.0cm
\centerline{\vbox{\epsfxsize=9.0cm\epsfbox{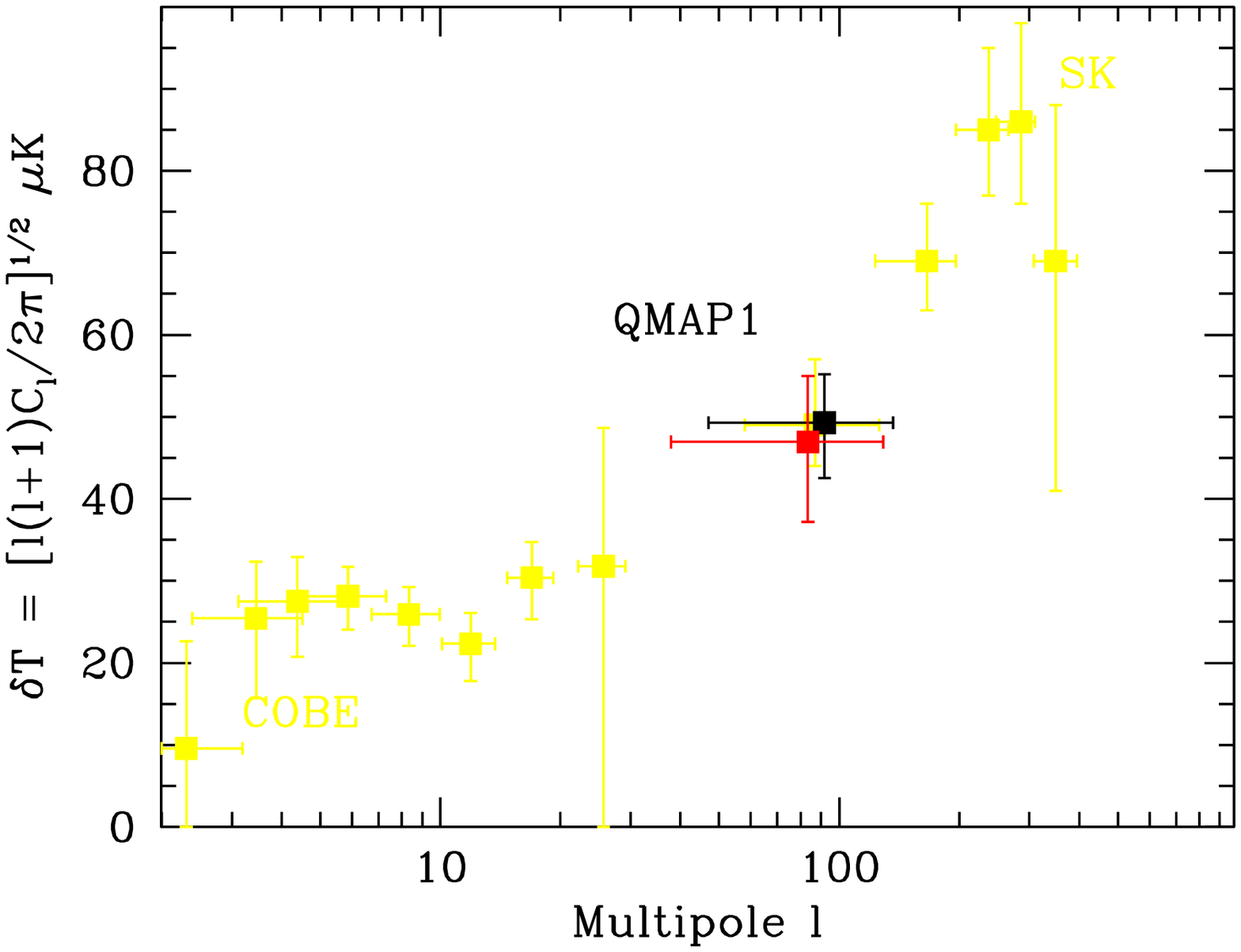}}}
\vskip-1.5cm 
\figcaption{
Angular power spectrum for flight 1.
The point on the left is for the Q-band and corresponds to
the band power 
$\delta T_\l = [\l(\l+1)C_\l/2\pi]^{1/2}=47^{+8}_{-10}$ $\mu$K 
in a window whose mean and 
rms width is given by $\l = 84\pm 46$. 
The point on the right is for the 
Ka-band and corresponds to $\delta T_\l = 49^{+6}_{-7}$ $\mu$K
in a window $\l = 92\pm 45$. These error bars do not include a 
calibration error of 12\% and 11\%, respectively.
}

%



 

\end{document}